\documentclass[preprint, 5p, twocolumn]{elsarticle}

\usepackage{lineno,hyperref}
\usepackage{lineno}
\usepackage{graphicx}
\usepackage{subcaption}
\usepackage{amsmath}
\usepackage{verbatim}
\usepackage{amssymb}
\usepackage{color}

\usepackage{mathrsfs}
\modulolinenumbers[5]

\journal{Journal of \LaTeX\ Templates}

\begin{document}

\begin{frontmatter}

\title{Kink solutions in generalized 2D dilaton gravity}


\author[address1]{Yuan Zhong\corref{mycorrespondingauthor}}
\cortext[mycorrespondingauthor]{Corresponding author}
\ead{zhongy@mail.xjtu.edu.cn}
\author[address2]{Heng Guo}
\author[address3]{Yu-Xiao Liu}
\address[address1]{MOE Key Laboratory for Nonequilibrium Synthesis and Modulation of Condensed Matter, \\ School of Physics, Xi’an Jiaotong University, Xi’an 710049, China}
\address[address2]{School of Physics and Optoelectronic Engineering, Xidian University, Xi’an 710071, China}
\address[address3]{Lanzhou Center for Theoretical Physics, Key Laboratory of Theoretical Physics of Gansu Province, and Key Laboratory of Quantum Theory and Applications of MoE, Lanzhou University, Lanzhou, Gansu 730000, China}

\begin{abstract}
We study static kink solutions in a generalized two-dimensional dilaton gravity model, where the  kinetic term of the dilaton is generalized to be an arbitrary function of the canonical one $\mathcal X= -\frac12 (\nabla \varphi)^2$, say $\mathcal F(\mathcal X)$, and the kink is generated by a canonical scalar matter field $\phi$. It is found that for arbitrary $\mathcal F(\mathcal X)$, the background field equations have a simple first-order formalism, and the linear perturbation equation can always be written as a Schr\"odinger-like equation with factorizable Hamiltonian operator.  After choosing appropriate $\mathcal F(\mathcal X)$ and superpotential, we obtain a sine-Gordon type kink solution with pure AdS$_2$ metric. The linear perturbation issue of this solution becomes an exactly solvable conformal quantum mechanics problem, if one of the model parameter takes a critical value.
\end{abstract}


\end{frontmatter}

%\linenumbers

%%%%%%%%%%%%%%%%%%%%%%%%%%%%%%%%%%%
%%%%%%%%%%%%%%%%%%%%%%%%%%%%%%%%%%%

%---------------------
\section{Introduction}

Two-dimensional (2D) gravity models allow physicists to study difficult issues like quantum gravity~\cite{Jackiw1985,Teitelboim1983,Henneaux1985,Alwis1992}, gravitational collapse~\cite{VazWitten1994,VazWitten1996}, black hole physics~\cite{BrownHenneauxTeitelboim1986,CallanGiddingsHarveyStrominger1992,BilalCallan1993,RussoSusskindThorlacius1992,RussoSusskindThorlacius1992a,RussoSusskindThorlacius1993,Ai2021}, and gauge/gravity duality~\cite{SachdevYe1993,Kitaev2015,AlmheiriPolchinski2015,MaldacenaStanfordYang2016,MaldacenaStanford2016,Jensen2016}, while avoiding technical complexity. For this reason, 2D gravity received much attention recently, see Refs.~\cite{Brown1988,Thorlacius1995,NojiriOdintsov2001d,GrumillerKummerVassilevich2002,Rosenhaus2019,Sarosi2018,Trunin2021,GrumillerRuzziconiZwikel2022} for comprehensive reviews.

Because the Einstein tensor vanishes for arbitrary 2D metric, one usually adopts the so-called dilaton gravity to describe 2D gravity, for example, the Jackiw-Teitelboim (JT) gravity~\cite{Jackiw1985,Teitelboim1983}:
\begin{equation}
S_{\rm{JT}}=\frac{1}{\kappa} \int d^{2} x \sqrt{-g} \varphi (R+\Lambda),
\end{equation}
where $\kappa$ and $\Lambda$ are the gravitational coupling constant and the cosmological constant, respectively. Obviously, in the JT gravity the dilaton field $\varphi$ plays the role of a Lagrangian multiplier, and the Ricci scalar $R$ is constrained to be a constant $-\Lambda$.

Another interesting 2D dilaton gravity is the one proposed by Mann, Morsink, Sikkema and Steele (MMSS), who added a kinetic term $\mathcal{X}\equiv -\frac{1}{2} g^{\mu\nu}\nabla_{\mu} \varphi \nabla_{\mu} \varphi$ to the action~\cite{MannMorsinkSikkemaSteele1991}:
\begin{equation}
S_{\mathrm{MMSS}}=\frac{1}{\kappa} \int d^2 x \sqrt{-g}(\varphi R+\mathcal{X}).
\end{equation}
In this case, the dilaton equation is
\begin{equation}
\label{eqDilaton0}
\nabla_\lambda \nabla^\lambda \varphi+R=0,
\end{equation}
which obviously allows solutions with variable scalar curvature. Since two-dimensional gravity has only one degree of freedom, it is always possible to express the metric as the following form~\cite{Stoetzel1995}:
\begin{equation}
\label{eqMetric0}
d s^2=-e^{2 A} d t^2+d x^2.
\end{equation}
One may notice that when the warp factor  $A=A(x)$ is static, which is assumed from now on, the above metric can be regarded as a 2D version of the Randall-Sundrum braneworld metric~\cite{RandallSundrum1999,RandallSundrum1999a}.

A remarkabe property of the MMSS gravity is that for the metric \eqref{eqMetric0} the dilaton equation \eqref{eqDilaton0} reduces to a simple algebraic relation~\cite{Stoetzel1995,Zhong2021}:
\begin{equation}
\varphi=2A.
\end{equation}
This relation enables us to eliminate $\varphi(x)$ in terms of $A(x)$, and therefore, largely reduces the complexity of the field equations. Especially, in some models with additional scalar matter fields, it was found that the field equations have very simple first-order formalisms, from which an important class of topological soliton solutions, namely, kink can be easily constructed~\cite{Stoetzel1995,Zhong2021,ZhongLiLiu2021,FengZhong2022,Zhong2022,LimaAlmeida2022,AndradeBazeiaLobaoJr.Menezes2022}.
Some of these kink solutions have asymptotic AdS$_2$ metrics, and can be interpreted as 2D thick branes. Besides, the linear perturbation equations of these solutions can always be rewritten as Schr\"odinger-like equations with factorizable Hamiltonians~\cite{Zhong2021,Zhong2021b}, which take similar forms as those of the scalar perturbations of 5D Einstein thick branes~\cite{Giovannini2003,Giovannini2001a,ZhongLiu2013}. The factorization of the perturbation Hamiltonian usually ensures the stability of the kink solutions~\cite{CooperKhareSukhatme1995}. If noncanonical scalar matter fields are allowed, it is even possible to construct 2D gravitating kink solutions with exactly solvable perturbation equations~\cite{Zhong2022}. As is well known, the information of linear spectrum plays a key role in understanding the quantum~\cite{ChristLee1975,Jackiw1977,Evslin2021,EvslinGuo2021,EvslinGarciaMartinCaro2022} and dynamic~\cite{CampbellSchonfeldWingate1983,DoreyMershRomanczukiewiczShnir2011,AdamOlesRomanczukiewiczWereszczynski2019a,MantonOlesRomanczukiewiczWereszczynski2021} properties of kink.

Since the MMSS gravity is just a special theory for 2D gravity, one may ask if there are other 2D dilaton gravity theories which have similar properties as the MMSS gravity in the modeling of gravitating kinks, namely:
\begin{enumerate}[(1)]
\item The field equations have simple first-order formalism, from which exact kink solutions can be easily constructed.
\item The Hamiltonian operator of the linear perturbation equation is factorizable.
\end{enumerate}

In this work, we report a 2D dilaton gravity model which extends the MMSS gravity but still reserves the above properties. The model and its general properties, including the first-order formalism and linear perturbation equations, are discussed in the next section. An explicit kink solution with pure AdS$_2$ metric will be derived in Sec.~\ref{Sec3}. The main results are summarized in Sec.~\ref{sec_con}.

%---------------------
\section{The model and its general properties }
\label{SecTwo}
We consider a generalized 2D dilaton gravity model with the following action
\begin{equation}\label{1}
S=\frac{1}{\kappa} \int d^{2} x \sqrt{-g}\left[\varphi R
+\mathcal F(\mathcal X)+\kappa \mathcal{L}_m\right],
\end{equation}
where $\mathcal{L}_m=-\frac12 (\nabla \phi)^2-V(\phi)$ is the Lagrangian density of the scalar matter field that generates the kink. What makes the present model different from the MMSS gravity is the term $\mathcal F(\mathcal X)$, which is an arbitrary function of the standard dilaton kinetic term $\mathcal X= -\frac12 (\nabla \varphi)^2$. The MMSS gravity model corresponds to the special case with $\mathcal F(\mathcal X)=\mathcal X$.

The action \eqref{1} leads to three field equations, namely, the dilaton equation
\begin{equation}
\label{STeq2}
\nabla^{\lambda} (\mathcal F_{\mathcal X}\nabla_{\lambda} \varphi)+R=0,
\end{equation}
the scalar equation
\begin{equation}
\label{STeq3}
 \nabla_{\lambda} \nabla^{\lambda} \phi=\frac{dV}{d\phi },
\end{equation}
and the Einstein equation
\begin{eqnarray}
\label{STeq1}
&&\mathcal F_{\mathcal X}\nabla_{\mu} \varphi \nabla_{\nu} \varphi-\frac{1}{2} g_{\mu \nu}\left(-2\mathcal F+4 \nabla_{\lambda} \nabla^{\lambda} \varphi\right)\nonumber\\
&+&2 \nabla_{\mu} \nabla_{\nu} \varphi+\kappa T_{\mu \nu}=0,
\end{eqnarray}
where $\mathcal{F}_{\mathcal X}$ and $\mathcal{F}_{\varphi}$ are the derivatives of $\mathcal{F}$ with respect to $\mathcal X$ and $\varphi$, respectively, and $T_{\mu \nu}=g_{\mu \nu}\mathcal{L}_m+ \nabla _{\mu}\phi \nabla _{\nu}\phi$ is the energy-momentum tensor.

For the static metric
\begin{equation}
\label{metricXCord}
  ds^2=-e^{2A(x)}dt^2+dx^2,
\end{equation}
 the dilaton and the scalar equations \eqref{STeq2} and \eqref{STeq3} become
\begin{equation}
\label{eqDilaton}
 \left( \mathcal{F}_{\mathcal{X}} \partial_x\varphi -2 \partial_x A\right)\partial_x A
 +\partial_x\left( \mathcal{F}_{\mathcal{X}} \partial_x\varphi -2 \partial_x A\right)=0,
\end{equation}
and
\begin{equation}
\label{eqScalar0}
\partial_x A\, \partial_x \phi+\partial_x^2 \phi=\frac{dV}{d\phi },
\end{equation}
respectively. The nontrivial components of the Einstein equation are
\begin{eqnarray}
\label{eqEin1}
&&-2 \partial_x^2 \varphi
-\partial_x \varphi (\mathcal{F}_{\mathcal{X}} \partial_x \varphi -2 \partial_x A
)=\kappa ( \partial_x \phi) ^2,
 \\
\label{eqEin2}
&&-2 \partial_x^2 \varphi+\mathcal{F}=  \frac12\kappa( \partial_x \phi) ^2+\kappa V.
\end{eqnarray}
Only three of the above four equations are independent. For example, one can derive  the scalar equation by using the dilaton equation and the Einstein equation. Thus, we will neglect the scalar equation \eqref{eqScalar0} and try to find solutions for the other three.
\subsection{The first-order formalism}
A remarkable feature of the present model is that the dynamical equations \eqref{eqDilaton}-\eqref{eqEin2} can be rewritten as a group of  first-order equations.

 To see this, we start by noticing that the dilaton equation \eqref{eqDilaton} is satisfied, if
\begin{equation}
\label{eqDilatonNonCon}
 \partial_x A= \frac12 \mathcal{F}_{\mathcal{X}} \partial_x\varphi,
\end{equation}
with which Eq.~\eqref{eqEin1}  reduces to
\begin{equation}
\label{eqEinFin1}
-2 \partial_x^2 \varphi=\kappa ( \partial_x \phi) ^2.
\end{equation}
To proceed, we introduce the so-called superpotential function $W(\phi)$ such that
\begin{equation}
\label{eqFirstPhi}
 \partial_x \phi= \frac{dW}{d\phi}.
\end{equation}
Then Eq.~\eqref{eqEinFin1} becomes
\begin{equation}
\label{eqFirstVarphi}
 \partial_x \varphi= -\frac\kappa 2 W,
\end{equation}
or equivalently, $\mathcal X(W)=-\frac{\kappa^2} 8W^2$. After substituting  Eqs.~\eqref{eqFirstPhi} and \eqref{eqFirstVarphi} into Eqs.~\eqref{eqEin2} and \eqref{eqDilatonNonCon} we obtain
\begin{eqnarray}
\label{eqFirstV}
V&=&\frac{\mathcal F}{\kappa }+\frac12\bigg(\frac{dW}{d\phi}\bigg)^2,\\
\label{eqFirstA}
 \partial_x A&=&- \frac\kappa4  \mathcal{F}_{\mathcal{X}}W.
\end{eqnarray}

As will be shown in Sec.~\ref{Sec3}, exact kink solutions can be easily constructed by inserting appropriate functions $W(\phi)$ and  $\mathcal{F}(\mathcal{X})$ into the first-order equations~\eqref{eqFirstPhi}-\eqref{eqFirstA}. We see that $\mathcal{F}(\mathcal{X})$ only affects the solutions of $V$ and $A$. Therefore, by modifying the function $\mathcal{F}(\mathcal{X})$, one can tune the form of the warp factor while keep $\phi$ and $\varphi$ unchanged.
\subsection{Linear stability issue}

It is convenient to discuss the linear perturbation in the conformally flat coordinates:
\begin{equation}
  ds^2=e^{2A(r)}(-dt^2+dr^2),
\end{equation}
where
\begin{equation}
\label{eqX2R}
r(x)\equiv\int_0^x e^{-A(\tilde x)}d\tilde x.
\end{equation}
For simplicity, we denote the derivatives with respect to $t$ and $r$ by overdots and primes, respectively.
In the $(t,r)$-coordinates the dynamical equations \eqref{eqDilaton}-\eqref{eqEin2} become
\begin{eqnarray}
\label{EqDilatonR}
A'&=&\frac{1}{2}\mathcal{F}_{\mathcal{X}} \varphi ', \\
\label{EqScaR}
\phi ''&=&e^{2 A}V_{\phi },\\
\label{EqEin1R}
V&=&\frac{1}{2} e^{-2 A} \phi '^2+\frac{\mathcal{F}}{\kappa },\\
\label{EqEin2R}
\varphi ''&=&\frac{1}{2}\mathcal{F}_{\mathcal{X}} \varphi '^2 -\frac{1}{2} \kappa  \phi '^2,
\end{eqnarray}
respectively.

Following Refs.~\cite{Zhong2021,ZhongLiLiu2021,Zhong2021b}, we consider small perturbations $\{\delta\varphi(r,t), \delta\phi(r,t), \delta g_{\mu\nu}(r,t)\}$ around an arbitrary static solution of Eqs.~\eqref{EqDilatonR}-\eqref{EqEin2R}, say, $\{\varphi(r), \phi(r), g_{\mu\nu}(r)\}$. It is convenient to define the metric perturbation as
\begin{eqnarray}
\delta g_{\mu\nu}(r,t)&\equiv& e^{2A(r)} h_{\mu\nu}(r,t)\nonumber\\
&=&e^{2A(r)} \left(
\begin{array}{cc}
 h_{00}(r,t) & \Phi (r,t) \\
 \Phi (r,t) & h_{rr}(r,t) \\
\end{array}
\right).
\end{eqnarray}
To the first order, the perturbation of the metric inverse reads
\begin{equation}
\delta g^{\mu \nu}=-e^{-2A} h^{\mu \nu},
\end{equation}
where
\begin{equation}
h^{\mu \nu}\equiv \eta^{\mu\rho}\eta^{\nu\sigma}h_{\rho\sigma}=\left(
\begin{array}{cc}
 h_{00} & -\Phi \\
 -\Phi & h_{rr} \\
\end{array}
\right).
\end{equation}
As  in Refs.~\cite{Zhong2021,ZhongLiLiu2021,Zhong2021b}, we define a variable $\Xi \equiv 2 \dot{\Phi}-h_{00}^{\prime}$, and taking the dilaton gauge $\delta \varphi=0$.

Independent perturbation equations can be obtained by linearizing the Einstein equation \eqref{STeq1} and the scalar field equation \eqref{STeq3}. The linearization of the Einstein equation leads to two independent perturbation equations, namely,
the $(0,1)$ component:
\begin{equation}
\label{eqPertOne}
{h}_{rr}=\kappa  \frac{ \phi '}{\varphi' } {\delta \phi } ,
\end{equation}
and the $(1,1)$ component:
\begin{equation}
\label{eqPertTwo}
\Xi =\kappa \frac{ \phi ' }{\varphi '} \bigg[\delta \phi '
+\delta \phi \left(\frac{ \varphi ''  }{\varphi '}
-\frac{   \phi ''}{\phi '} - \mathcal{F}_{\mathcal{X}\mathcal{X}}\mathcal{X} \varphi ' \right)\bigg] .
\end{equation}
The $(0,0)$ component is also nontrivial, but after substituting the background field equations it reduces to Eq.~\eqref{eqPertOne}.

Another independent perturbation equation comes from the linearization of the scalar equation \eqref{STeq3}, which, after eliminating $h_{rr}$ and $\Xi$ by using Eqs. \eqref{eqPertOne} and \eqref{eqPertTwo}, takes the following form:
\begin{eqnarray}
\label{eq50}
&&\ddot{\delta \phi } -\delta \phi ''-\bigg[\mathcal{F}_{\mathcal{XX}}\mathcal{X}\left( \varphi ''-\frac{1}{2}\mathcal{F}_{\mathcal{X}} \varphi ^{\prime 2}\right)\nonumber\\
& +&\mathcal{F}_{\mathcal{X}}\left( \varphi ''-\frac{\phi ''}{\phi '} \varphi '\right)
 -2\left(\frac{\varphi ''}{\varphi '}\right)^{2} -\frac{\phi '''}{\phi '} \nonumber\\
& +&4\frac{\varphi  ''}{\varphi '}\frac{\phi ''}{\phi '}\bigg] \delta\phi=0.
\end{eqnarray}
The terms that contain $\mathcal{F}_{\mathcal{X}}$ and $\mathcal{F}_{\mathcal{X}\mathcal{X}}$ can be eliminated by applying the following identity:
\begin{eqnarray}
&&\mathcal F_{\mathcal{XX}} \mathcal X\left( \varphi ''-\frac{1}{2}\mathcal{F}_{\mathcal{X}} \ \varphi ^{\prime 2}\right)
\nonumber\\
&+&\mathcal{F}_{\mathcal{X}}\left( \varphi ''-\frac{\phi ''}{\phi '} \varphi '\right)=\frac{\varphi '''}{\varphi '} -2\frac{\varphi  ''}{\varphi '}\frac{\phi ''}{\phi '},
\end{eqnarray}
which is derived by using background equations \eqref{EqDilatonR} and \eqref{EqEin2R}.
Finally, the equation for $\delta\phi$ takes the same form as the one derived in the MMSS gravity~\cite{Zhong2021}:
\begin{equation}
\label{eqPertFinal}
\ddot{\delta \phi } -\delta \phi ''+\left[\frac{\phi '''}{\phi '} -2\frac{\varphi ''\phi ''}{\varphi '\phi '} +2\left(\frac{\varphi ''}{\varphi '}\right)^{2} -
\frac{\varphi '''}{\varphi '}\right]\delta\phi =0,
\end{equation}
which can also be written as
\begin{equation}
\ddot{\delta \phi } -\delta \phi ''+
\frac{f''}{f}\delta\phi =0,\quad f= \frac{\phi'}{\varphi'}.
\end{equation}

By conducting the mode expansion
\begin{equation}
\label{eqMode}
\delta\phi=\sum_n \psi_n(r)e^{i \omega_n t},
\end{equation}
we can rewrite the perturbation equation into a Schr\"odinger-like equation:
\begin{equation}
\label{EqSchR}
\hat H \psi_n\equiv \left[-\frac{d^2}{dr^2}+V_{\textrm{eff}}\right]\psi_n=\omega_n^2 \psi,
\end{equation}
where the effective potential $V_{\text{eff}}\equiv \frac{f''}{f}$.  For continuous eigenvalues, the summation in Eq.~\eqref{eqMode} should be understood as integration. The particular form of the effective potential enables us to factorize the Hamiltonian operator into the product of an operator $\hat{ \mathcal{A}}$ and its Hermit conjugate:
\begin{equation}
\hat H ={ \hat {\mathcal{A}}^\dagger}{ \hat{\mathcal{A}}},
\end{equation}
where
\begin{equation}
\hat{\mathcal{A}}=-\frac{d}{d r}+\frac{f'}{f}, \quad
\hat{\mathcal{A}}^\dagger=\frac{d}{d r}+\frac{f'}{f}.
 \end{equation}

According to the theory of supersymmetric quantum mechanics~\cite{CooperKhareSukhatme1995},   the eigenvalues of such a factorizable Hamiltonian operator are semipositive definite, namely, $\omega_n^2\geq 0$. Therefore, any static solution is stable against small linear perturbations. The ground state has vanished eigenvalue $\omega_0= 0$, and the corresponding wave function is $\psi_0(r) \propto f$. 

Now, let us consider an explicit kink solution.
%=============================================
\section{Kink with AdS$_2$ metric}
\label{Sec3}
To construct an explicit solution, one must specify the functions $\mathcal F(\mathcal X)$ and $W(\phi)$.
As can be seen from Eq.~\eqref{eqDilatonNonCon}, the freedom in choosing $\mathcal F(\mathcal X)$ allows us  to construct kink solutions with very simple warp factor. For example, if we take
\begin{equation}
\mathcal F=-2 \sqrt{-2\mathcal X}/l,
\end{equation}
such that $\mathcal F_{\mathcal X}=\frac {2} {l\,|\partial_x \varphi |}$, then Eq.~\eqref{eqDilatonNonCon} leads to a metric solution of the following form:
\begin{equation}
A(x)=\textrm{sgn}(\partial_x \varphi)\cdot  x/l,
\end{equation}
where $l>0$ is a parameter with the dimension of length, and $\textrm{sgn}(x)$ is the sign function.
Obviously, if we can construct  solutions with monotonically increasing dilatons, such that $\textrm{sgn}(\partial_x \varphi)=1$, then the warp factor
\begin{equation}
\label{eqMetric}
A(x)= x/l
\end{equation}
describes a pure AdS$_2$ space with negative constant curvature $R=-2 l^{-2}$.

As can be seen from Eqs.~\eqref{eqFirstPhi} and \eqref{eqFirstVarphi}, the solutions of $\varphi$ and  $\phi$ are completely determined by the superpotential $W(\phi)$. Therefore, by choosing suitable superpotentials we can obtain monotonically increasing dilatons, one such example is~\cite{OmotaniSaffinLouko2011}
\begin{equation}
W(\phi)=k v^2 \left[\sin \left(\frac{\phi}{v}\right)-c\right],
\end{equation}
for which Eqs.~\eqref{eqFirstPhi}-\eqref{eqFirstV} have the following solution:
\begin{eqnarray}
\label{solPhi}
&& \phi(x)=v \arcsin (\tanh (k x)), \\
\label{solVarphi}
&& \varphi(x)=\frac{1}{2} \kappa  v^2 [c k x-\ln (\cosh (k x))],\\
&&V(\phi)=\frac{1}{2} \left[\cos ^2\phi -2 l (c-\sin \phi)\right],
\end{eqnarray}
where $k, v, c$ are some real parameters.

\begin{figure*}[!ht]
\centering
\includegraphics[width=1\textwidth]{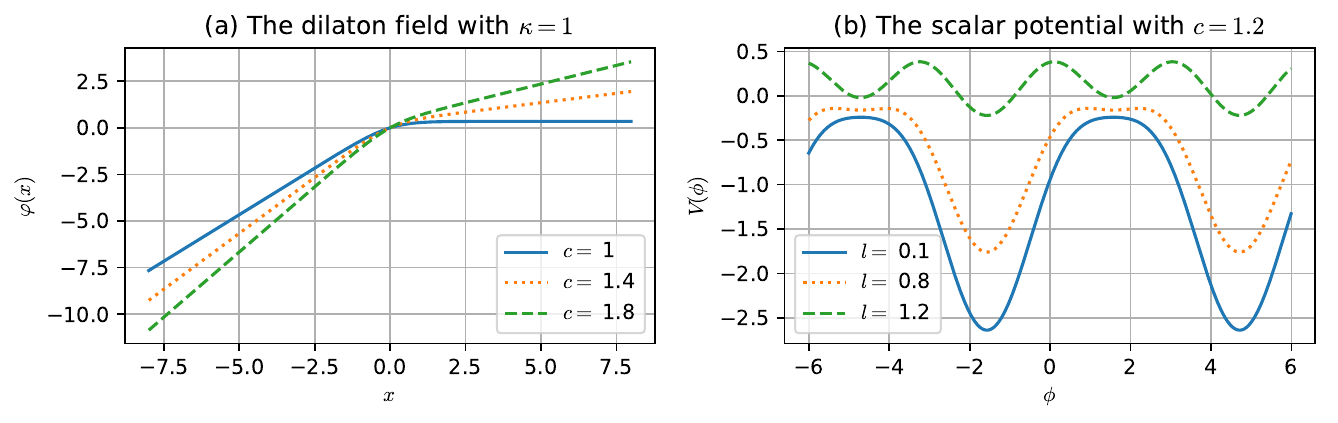}
\caption{Plots of (a) the dilaton field $\varphi(x)$, and (b) the scalar potential $V(\phi)$. For $c\geq 1$, the dilaton field is monotonically increasing. }
\label{Fig1}
\end{figure*}

The scalar field configuration in Eq.~\eqref{solPhi} corresponds to a sine-Gordon kink, whose  width and the asymptotic behavior are controlled by parameters $k$ and $v$, respectively. For simplicity, we fix $k=v=1$, so that $\lim_{x\to\pm\infty}\phi(x)=\pm \pi/2$. For the dilaton field, the asymptotic behavior is
\begin{equation}
\lim_{x\to\pm\infty}\varphi(x)=\frac{1}{2} \kappa  [(c\mp1) x+\ln 2].
\end{equation}
Obviously, as the dimensionless parameter $c$ is turned on, the dilaton becomes asymmetric.  Especially, for  $c\geq 1$, $\partial_x \varphi=\frac{1}{2} \kappa [c-\tanh ( x)] \geq0$, and the dilaton becomes a monotonically increasing function, see Fig.~\ref{Fig1} (a). In the critical case $c=1$, the dilaton approaches to a constant $\frac{1}{2} \kappa \ln 2$, as $x\to+\infty$.

In what follows, we assume $c\geq 1$ so that the metric solution is the one given in \eqref{eqMetric}, and the integral \eqref{eqX2R} gives
\begin{equation}
r=(1-e^{-  x/l})l.
\end{equation}
It is convenient to introduce two dimensionless variables $\alpha\equiv k\,l>0$ and $u\equiv e^{ -x/l}\in [0,+\infty)$\footnote{Note that $x=-\infty,0,+\infty$ is mapped to $u=+\infty, 1, 0$, respectively.
}, in terms of which the scalar and dilaton fields read
\begin{eqnarray}
\label{phiu}
&& \phi(u)=-\arcsin[\tanh (\alpha \ln u)],\\
\label{varphiu}
&& \varphi(u)=-\frac{1}{2} \kappa  v^2 [\alpha  c \ln u+\ln [\cosh (\alpha  \ln u)]].
\end{eqnarray}
Using the relation $\frac{d}{dr}=\frac{du}{dr}\frac{d}{du}=-l^{-1} \frac{d}{du}$, the Schr\"odinger-like equation \eqref{EqSchR} can be rewritten as
\begin{equation}
\label{EqSchU}
-\frac{d^2}{du^2}\psi_n+\tilde V_{\textrm{eff}}\, \psi_n=\tilde{w}^2_n \psi_n,
\end{equation}
where $\tilde V_{\textrm{eff}}=\partial_u^2 f/f$ and $\tilde{w}_n= {w_n}{l}$. A direct calculation gives
\begin{eqnarray}
\tilde V_{\textrm{eff}}(u)&=&\alpha u^{-2} \left[(c+1) u^{2 \alpha }+c-1\right]^{-2} \nonumber\\
&\times&\big[-6 \alpha  \left(c^2-1\right) u^{2 \alpha }+(\alpha -1) (c-1)^2\nonumber\\
&+&(\alpha +1) (c+1)^2 u^{4 \alpha }\big],
\end{eqnarray}
and
\begin{eqnarray}
\label{eqfu}
 \psi_0(u)\propto f(u)=\frac{4 u^{\alpha }}{\kappa   \left[(c+1) u^{2 \alpha }+c-1\right]}.
\end{eqnarray}

\begin{figure*}[!ht]
\centering
\includegraphics[width=1\textwidth]{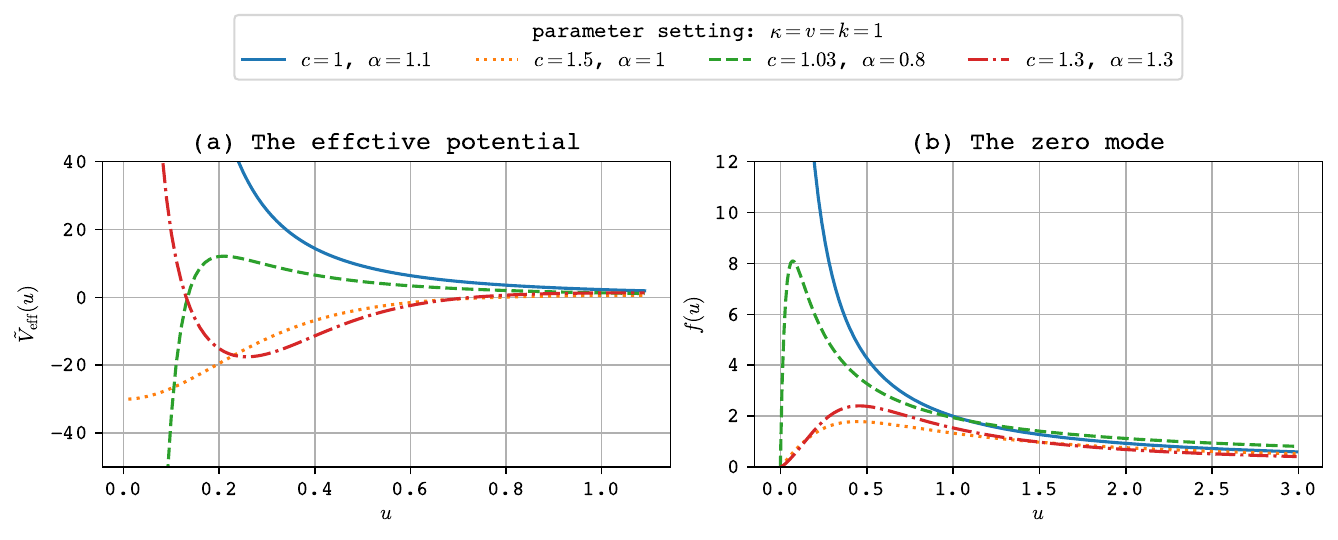}
\caption{Plots of (a) the effective potential $\tilde V_{\textrm{eff}}(u)$, and (b) the zero mode $\psi_0(u)\propto f(u)$.  }
\label{Fig2}
\end{figure*}

Obviously, as $u\to +\infty$,  $\tilde V_{\textrm{eff}}\sim \frac{\alpha  (\alpha +1) }{u^2}$ approaches to zero, for examples see Fig.~\ref{Fig2} (a). Thus, there is no bounded shape mode in the spectrum, and all the eigenstates with positive eigenvalues form a continuum. However, there might be a normalizable zero mode, depending on the behavior of $\tilde V_{\textrm{eff}}(u)$ around $u=0$, and there are four different cases.
\subsection*{ Case 1: $c=1$.}
In this critical case
\begin{equation}
\label{CQM}
\tilde V_{\textrm{eff}}(u)=\frac{\alpha(\alpha+1)}{u^2},
\end{equation}
is positive for $\alpha>0$. For this simple potential, Eq.~\eqref{EqSchU} is exactly solvable, and the regular solution takes the following form:
\begin{equation}
\psi_n(u)\propto \sqrt{u} \, J_{\alpha+\frac{1}{2}}(\tilde w_n u), \quad \tilde w_n>0,
\end{equation}
where $J_m(x)$ is the spherical Bessel function of order $m$. The zero mode wave function
\begin{equation}
\psi_0\propto f \propto u^{- \alpha }
\end{equation}
is divergent at $u=0$, and therefore, cannot be normalized.

\subsection*{ Case 2: $c>1$, $\alpha= 1$.}
In this case,
\begin{equation}
\tilde V_{\textrm{eff}}(u)=
\frac{2 (c+1) \left[ \left( c+1\right)u^2-3(c-1)\right]}{\left[(c+1) u^2+c-1\right]^2}.
\end{equation}
Unlike case 1 where $\tilde V_{\textrm{eff}}$ is divergent and positive at $u=0$, here
\begin{equation}
\tilde V_{\textrm{eff}}(0)=
-\frac{6 (c+1)}{c-1}
\end{equation} is  finite and negative, see Fig.~\ref{Fig2} (a). The zero mode
\begin{equation}
\psi_0= \mathcal{N} \frac{ u}{(c+1) u^2+c-1} \propto f,
\end{equation}
is nodeless and normalizable, and the normalization constant  $\mathcal{N}=\left[\frac{4 (c+1) \sqrt{c^2-1}}{\pi }\right]^{1/2}$ is finite. 

For $c>1, \alpha\neq1$, the effective potential behaves as $\frac{(\alpha -1) \alpha }{u^2}$ when $u\to 0$.
Therefore, there are  two other cases with singular effective potentials.
\subsection*{ Case 3: $c>1$, $0<\alpha< 1$.}
In this case, we get an attractive volcano-like potential, and the zero mode is normalizable if $\alpha>\frac12.$  For $\alpha\leq \frac12$, the zero mode wave function is regular, but not square integrable.

\subsection*{ Case 4: $c>1$, $\alpha> 1$.}
In this case, the effective potential has a repulsive core at $u=0$, and a finite well within $u\in(0,1)$. The zero mode is always nodeless and normalizable.

To summarize, if $c>1, \alpha>1/2$, the zero mode has a normalizable wave function, which has no node, and therefore, is the ground state of the system. As can be seen from Eq.~\eqref{eqfu}, the peak of the zero mode $f(u)$ locates at $u=u_{\textrm{max}}= \left(\frac{c-1}{{c+1}}\right)^{1/2\alpha }$, where $f(u_{\textrm{max}})=\frac{2}{\sqrt{c^2-1}}$ is independent of $\alpha$. For $c\geq 1$, we always have $u_{\textrm{max}}<1$, and if $c\gg 1$, we have $u_{\textrm{max}}\approx 1$. Plots of the effective potential and the zero mode can be found in Fig.~\ref{Fig2}.

Particular attention should be paid, however, to the critical case with $c=1$, for which the effective potential reduces to an inverse-square potential $\tilde V_{\textrm{eff}}(u)=\frac{\alpha(\alpha+1)}{u^2}
$. With this potential, the Schr\"odinger-like equation \eqref{EqSchU} is not only exactly solvable, but also conformally invariant~\cite{AlfaroFubiniFurlan1976}. Therefore, we have revealed an interesting relation between  AdS kink solutions and conformal quantum mechanics. In fact, similar relations were also found in other models. 

For example, in Liouville model, the nonlinear field equation supports a static space-dependent  solution, for which the linear perturbation modes obey a Schr\"odinger-like equation with the aforementioned inverse-square potential with $\alpha=1$~\cite{DHokerFreedmanJackiw1983,DHokerJackiw1983}. 
As explained in Refs.~\cite{DHokerFreedmanJackiw1983,DHokerJackiw1983},  the absence of the zero mode in Liouville model, is a consequence of the spontaneous breaking of the  space-translation symmetry. Without the normalizable zero mode, there is no infrared divergences in the corresponding quantum theory, and the propagator can be  explicitly constructed. Finally, the perturbation theory is finite after ultraviolet mass renormalization~\cite{DHokerFreedmanJackiw1983,DHokerJackiw1983}. 

In fact, static solutions without normalizable zero mode is not only acceptable physically, but also required in some cases. For example, in most 5D thick brane models, the zero mode of the scalar perturbation must not be normalizable, otherwise, one would confront with the problem of fifth force~\cite{Giovannini2003,Giovannini2001a,ZhongLiu2013}.

\section{Conclusion}
\label{sec_con}
In this work, we studied a 2D dilaton gravity where the dilaton has a generalized kinetic term $\mathcal F(\mathcal X)$.  We found that the static field equations have a simple first-order formalism, from which exact kink solutions can be easily constructed after giving suitable superpotential $W(\phi)$ and function $\mathcal F(\mathcal X)$. The solutions of the dilaton and the scalar fields are determined only by the form of $W(\phi)$. While the solution of the metric depends on both $W(\phi)$ and $\mathcal F(\mathcal X)$.  Therefore, by tuning $\mathcal F(\mathcal X)$, one may obtain kink solutions with different metrics. The example we given here is a sine-Gordon kink with a pure AdS$_2$ metric, but it is not difficult to construct many other solutions.

We also found that for arbitrary static solutions of our model, the linear perturbation equation can always be written as a Schr\"odinger-like equation with factorizable Hamiltonian operator.  The factorization of the Hamiltonian not only ensures the semi-positivity of the spectrum (and therefore the stability of the solution), but also gives the analytical expression of the zero mode wave function. 

In particular, for our AdS$_2$ kink solution, there is no bounded shape modes in the linear spectrum, because the effective potential approaches to zero as $u\to +\infty$. But,  if $c>1$ and $\alpha>1/2$ there is a normalizable zero mode as the ground state. While, for $c=1$ the linear perturbation issue becomes a conformal quantum mechanics problem, and the effective potential is an exactly solvable inverse-square potential without normalizable zero mode.

It is interesting to explore other kink solutions by choosing different $W(\phi)$ and $\mathcal F(\mathcal X)$, or  to investigate the quasi-normal mode issues of gravitating kinks by following the discussions of Refs.~\cite{TanGuoLiu2022,TanGuoZhangLiu2023}. We leave these issues to our future works.

\section*{Acknowledgments}
This work was supported by the National Natural Science Foundation of China (Grant numbers 12175169, 12247101 and 11875151), and the Natural Science Basic Research Plan in Shaanxi Province of China (Program No. 2020JM-198).

\section*{Declaration of competing interest}
The authors declare that they have no known competing financial interests or personal relationships that could have appeared to influence the work reported in this paper.

%%%%%%%%%%%%%%%%%%%%%%%%%%%%%%%%%%%
%%%%%%%%%%%%%%%%%%%%%%%%%%%%%%%%%%%
%\section*{Bibliography}
%%%%%%%%%%%%%%%%%%%%%%%%%%%%%%%%%%%
%%%%%%%%%%%%%%%%%%%%%%%%%%%%%%%%%%%


%\bibliography{/Users/zhongy/360Yun/jabref/library/articles}%macbook
%\bibliography{/Users/zhongyuan/360Yun/jabref/library/articles}% macstudio
%\bibliography{E:/360YunPan/jabref/library/articles}% Produces the bibliography via BibTeX.

\end{document}